\documentstyle[preprint,aps]{revtex} 
\input{epsf.sty}
\tightenlines
\begin{document}

\draft

\title{On the $\overline{d}/\overline{u}$ Asymmetry and Parton Distributions.} 
\author{A. Daleo,$^{1}$
C.A. Garc\'{\i}a Canal,$^1$ G.A. Navarro,$^2$ R. Sassot$^2$}
\address{$^1$  Laboratorio de F\'{\i}sica Te\'{o}rica,
 Departamento de F\'{\i}sica,
Universidad Nacional de La Plata,
 C.C. 67, 1900 La Plata, Argentina \\
 $2$ Departamento de F\'{\i}sica, Universidad de Buenos Aires,
 Ciudad Universitaria, Pab.1, 1428 Buenos Aires, Argentina}
\maketitle

\begin{abstract}
We discuss the impact of different measurements of the $\overline{d}/\overline{u}$ asymmetry in the extraction of parameterizations of parton distribution functions.
\end{abstract}

\section{Introduction}

Recently, the E866/NuSea Collaboration \cite{E866n} has released the final results corresponding to the analysis of their full data set of Drell-Yan yields from an 800 GeV/c proton beam on hydrogen and deuterium targets. This analysis, aimed to size the $\overline{d}/\overline{u}$ asymmetry in the proton, has confirmed previous measurements by E866 \cite{E866}, but with a substantial increase in precision and extending the kinematical coverage.  

The measurement of the $\overline{d}/\overline{u}$ asymmetry in the proton has
been pursued in recent years by several groups \cite{NMC94,NMC97,NA51,Hermes,E772}, triggered by the NMC Collaboration measurement of deep inelastic muon scattering from hydrogen and deuterium targets, which suggested that the $\overline{d}$ quark content in the proton
was larger than that of the $\overline{u}$ quarks \cite{NMC}. 

The ongoing interest in this issue has been not only the verification a hitherto unexpected flavour symmetry breakdown, presumably of non-perturbative origin, but also the precise determination of the ratio $\overline{d}/\overline{u}$, as required by modern global QCD fits of parton distribution functions \cite{GRV,MRST,CTEQ}. These global analyses are an essential ingredient in the perturbative QCD  description of hadrons and their hard interactions, and in recent years have evolved in accuracy and sophistication along the increase in precision of data \cite{IJMP}.   

Indeed, prior to NMC finding, global extractions of parton distributions were
made under the assumption of flavour symmetry for the light sea quarks ($\overline{u}=\overline{d}$). Most recent sets, however, were constrained to satisfy either NMC or, more likely, E866 data, imposition that also influences the extraction of valence quark densities \cite{GRV,MRST,CTEQ}. 

It is interesting to notice at this point, that the constraints on sea quark densities coming from the various asymmetry measurements are not 
necessarily equivalent and, eventually,  they influence in different 
ways the extraction of parton densities in global fits. These differences are most noticeable when comparing for example E866 Drell Yan data and NMC results on the difference between the proton and the neutron structure functions. In other words, global fits designed to accommodate E866 data fail to reproduce that coming from  NMC, and conversely.

Given the accuracy of modern parton distributions, and the fact that $F_2^p$ and $F_2^d$, as well as E866 data are nicely accounted for by them, it is somewhat surprising that they fail to reproduce $F_2^p-F_2^n$ data at intermediate values of the variable $x$, as shown in Figure 1.  

\firstfigfalse
\setlength{\unitlength}{1.mm}
\begin{figure}[!hbt]
\begin{picture}(100,175)(0,0)
\put(-15,55){\mbox{\epsfxsize11.0cm\epsffile{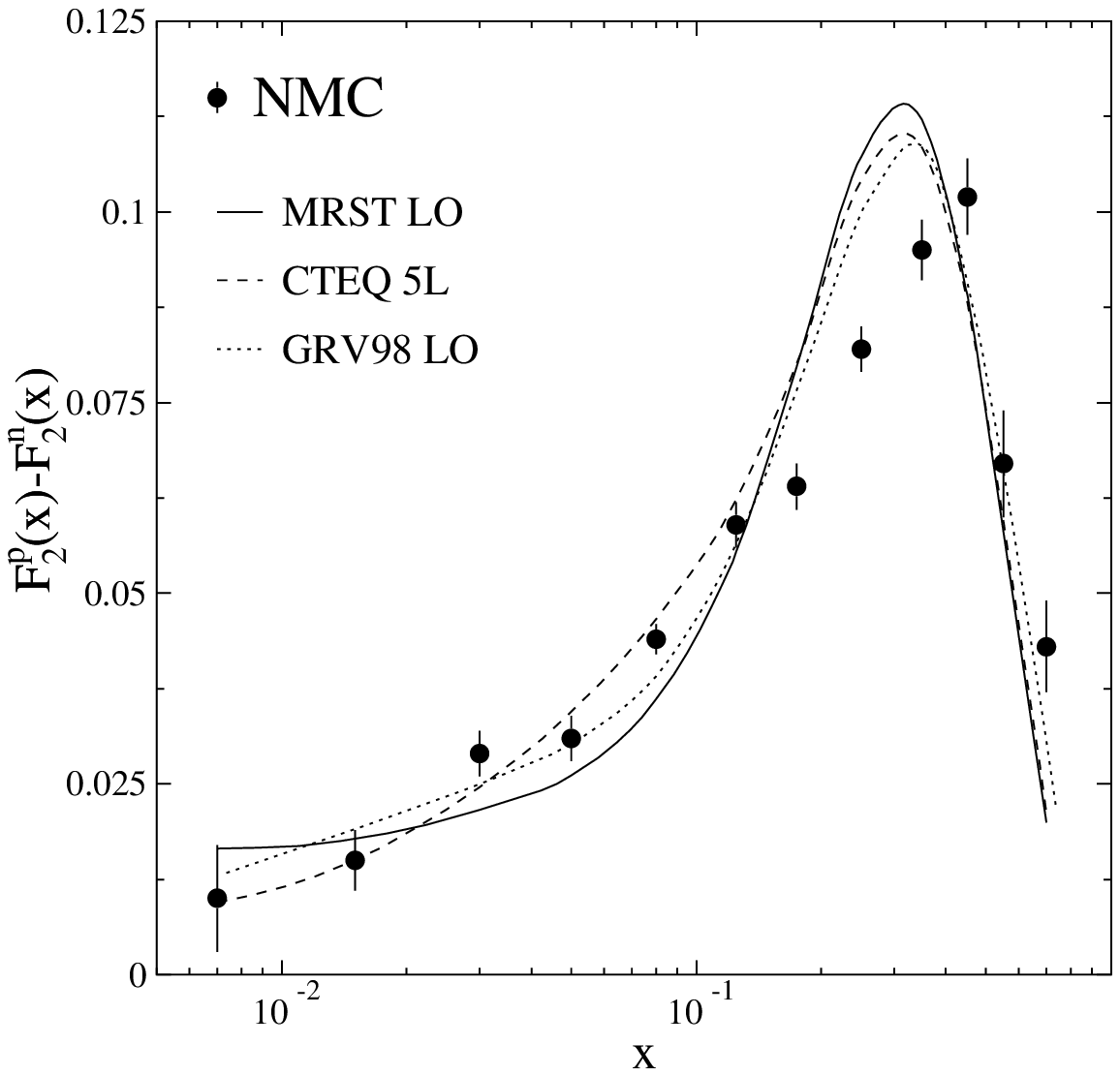}}}
\put(68,55){\mbox{\epsfxsize11.0cm\epsffile{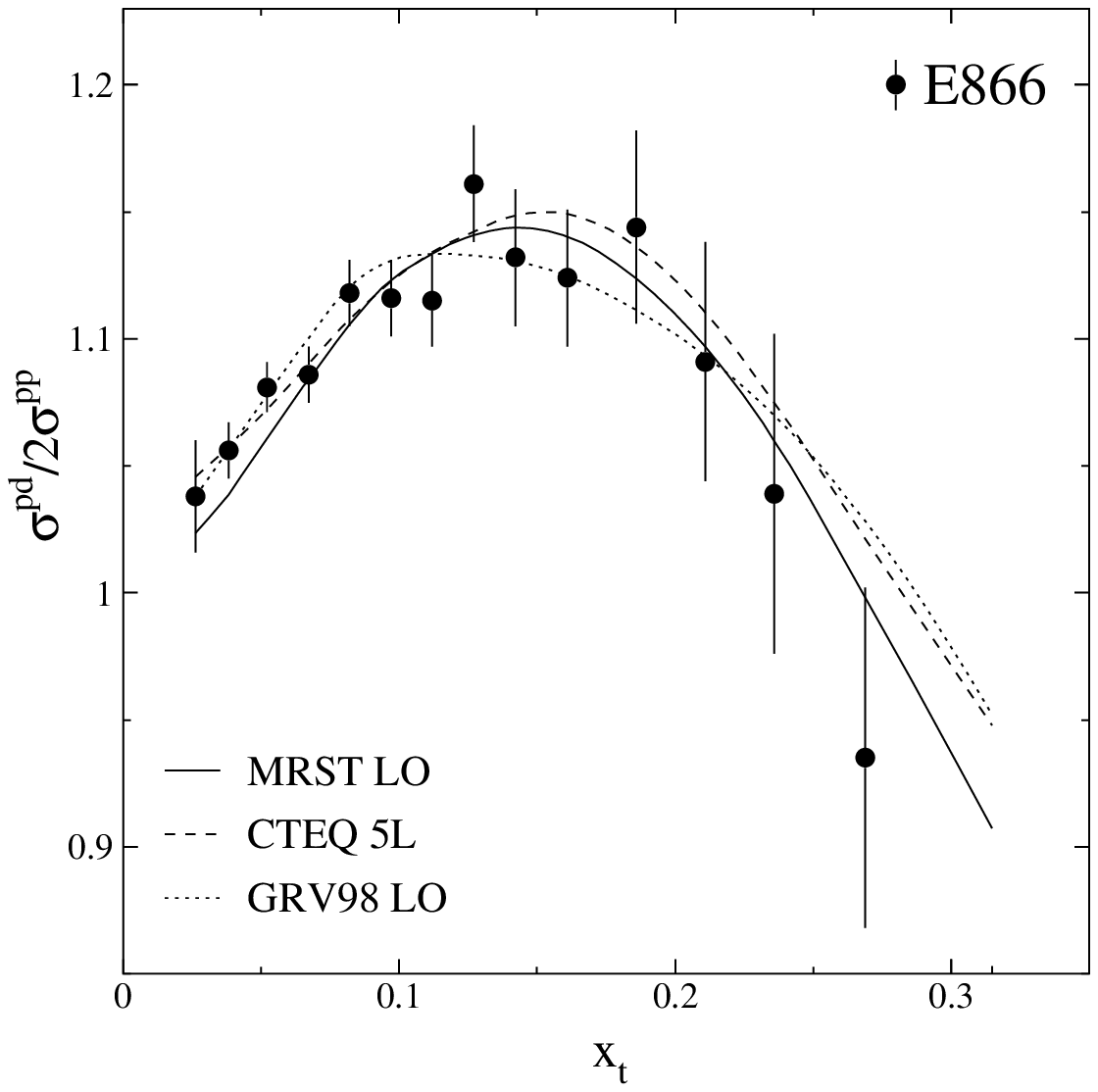}}}
\put(30,-27){\mbox{\epsfxsize11.0cm\epsffile{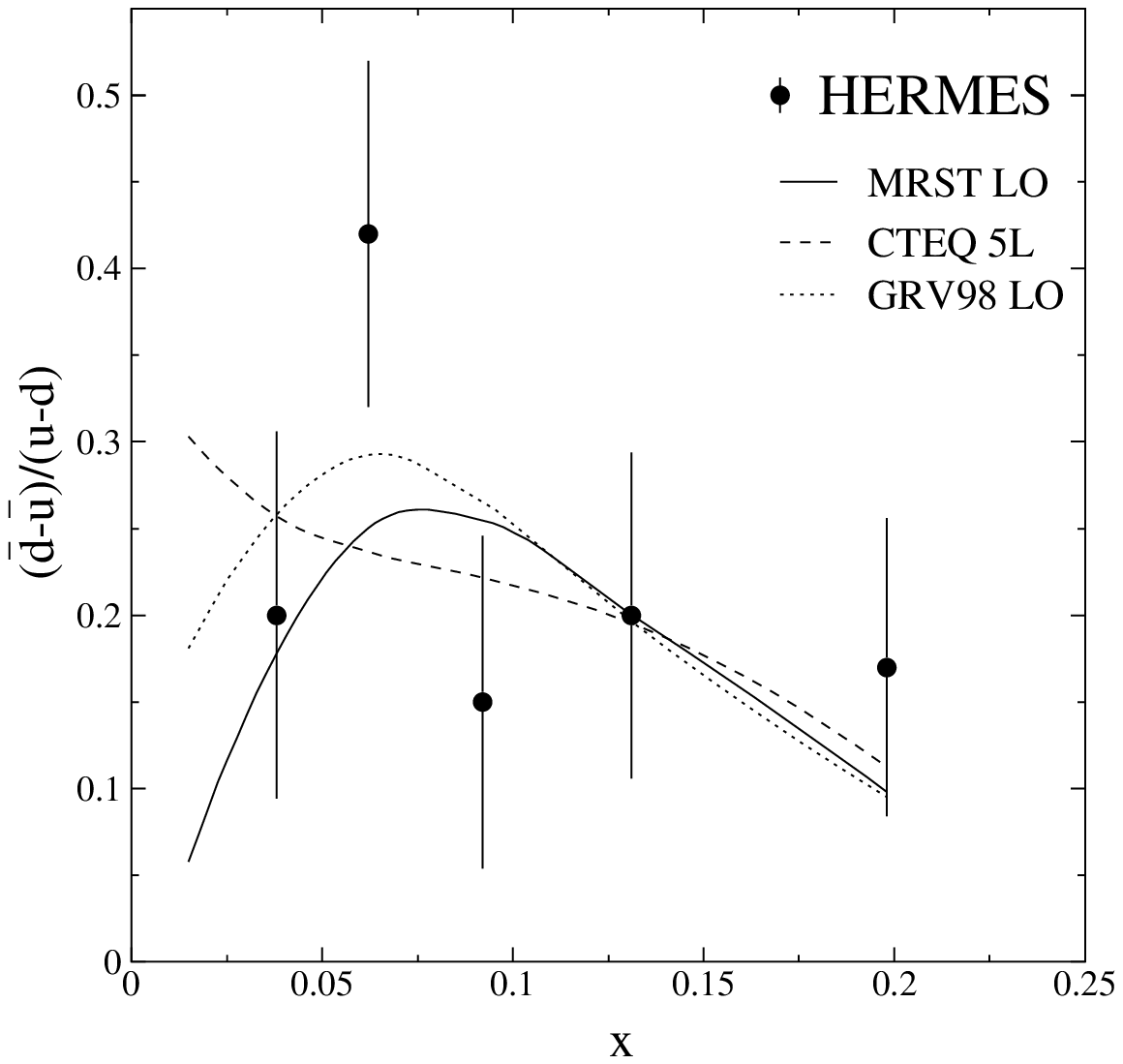}}}
\put(65,165){\mbox{\footnotesize \bf a)}}
\put(91,165){\mbox{\footnotesize \bf b)}}
\put(53,82){\mbox{\footnotesize \bf c)}}
\put(5,5){\mbox{\footnotesize \bf FIGURE 1:} 
{\footnotesize NMC (a), E886 (b), and HERMES (c) $\overline{d}/\overline{u}$-related measurements against current }} 
\put(22,0){\mbox{{\footnotesize LO parameterizations.
  }}}
\end{picture}
\end{figure}

The failure in reproducing the different data sets simultaneously may be pointing out the need of more precise data, the necessity of a revision of some of the underlying assumptions in the analysis of the data, such as the role of nuclear corrections in the use of deuterium targets ~\cite{Melni,Bodek,Rolo}, for example, or even the presence of some other hitherto neglected effect.  In the mean time, it is worthwhile analysing to what extent parton distributions are affected by the different alternatives.
  
In order to assess the impact of different results on the $\overline{d}/\overline{u}$ asymmetry in the extraction of parton distribution functions, we  perform in this paper a leading order (LO) QCD global extraction of parton densities from a large set of DIS data as usual, but forcing the agreement with one particular flavour asymmetry measurement (NMC, E866,  or HERMES) as an {\em ad hoc} constraint in the fit. This is accomplished giving to the `preferred' data set a larger statistical 
weight than it actually has, and it is done also in order to increase the little relative statistical relevance that they have against the bulk of DIS data.

As result of this approach we find that even though one can produce parton densities using either E866 or NMC data to implement the flavour symmetry breaking, the constraints that these sets of data impose
on the distributions can not be satisfied simultaneously in any way. HERMES data show a similar conflict with NMC, but is in rather good agreement with E866.

Parton distributions in agreement with E866/HERMES symmetry breaking scenario lead in the fits to lower global $\chi^2$ values than those that reproduce NMC data, and
the shape of the sea  and valence quark densities of these extreme
scenarios show clearly noticeable differences. 

In the following section we summarize the main features of the parton distributions parameterizations used, the $Q^2$-evolution strategy, and the data included in the global analyses. Then we present the outcome of the three fits
in which either E866, NMC, or HERMES data have been favored, against the standard one in which no additional weights have been applied. Finally, we present our conclusions.

\section{Parton Densities and Data.}

In order to parameterize quark densities at the initial scale $Q^2_0$ we adopt the
standard functional dependence as in ~\cite{MRST}:
\begin{eqnarray}
x\,u_v(x,Q^2_0) &=& N_u\, x^{\alpha_u} (1-x)^{\beta_u} (1+\gamma_u \sqrt{x}+\delta_u x) \nonumber \\
x\,d_v(x,Q^2_0) &=& N_d\, x^{\alpha_d} (1-x)^{\beta_d} (1+\gamma_d \sqrt{x}+\delta_d x)
\end{eqnarray}
for valence quarks;
\begin{eqnarray}
x\,\Sigma(x,Q^2_0) = N_{\Sigma}\, x^{\alpha_{\Sigma}} (1-x)^{\beta_{\Sigma}}
 (1+\gamma_{\Sigma} \sqrt{x}+\delta_{\Sigma} x)
\end{eqnarray}
for the sum of light sea quarks and antiquarks ($u_s+\overline{u}_s+d_s+\overline{d}_s+s_s+
\overline{s}_s$) and
\begin{equation}
xg(x,Q^2_0)=N_g\, x^{\alpha_g} (1-x)^{\beta_g} (1+\gamma_g \sqrt{x}+\delta_g x)
\end{equation}
for gluons. In order to be consistent with CCFR data on dimuon
production~\cite{CCFR}, strange sea quarks are assumend to be
 $\overline{s} = 0.1\, \Sigma$ while $\overline{u}$ and
$\overline{d}$ densities are given by
\begin{eqnarray}
x(\overline{d}+\overline{u})& = & 0.4 \, {\Sigma} \nonumber \\
x(\overline{d}-\overline{u})& = & N_{\Delta}\, x^{\alpha_{\Delta}}
(1-x)^{\beta_{\Delta}}
 (1+\gamma_{\Delta} \sqrt{x}+\delta_{\Delta} x^2)
\end{eqnarray}
Leaving aside the normalizations for valence quarks and gluons,
which are fixed by charge and momentum conservation, the above
distributions imply 21 free parameters to fit.

The initial scale $Q^2_0$ is taken to be $1 \,\mbox{GeV}^2$ and the
evolution is performed using a va\-ria\-ble number of active flavours
with thresholds at $1.69 \,\mbox{GeV}^2$  for charm, and
 $18.49\,\mbox{GeV}^2 $ for bottom. Bellow these thresholds the contributions to the
 structure function are given  by the corresponding
 photon-gluon fusion diagrams.
 For the strong coupling $\alpha_s$
 the LO expression with $\Lambda_{QCD}=0.174\,\mbox{GeV}$ (for four flavours) is taken.

The data set used in the global fitting procedure is listed in Table 1, where we have excluded data with $x >0.75$, $x < 0.001$,  $Q^2 < 2
\,\mbox{GeV}^2$, and $W^2 < 7\, \mbox{GeV}^2$ in order to avoid higher order QCD
and mass effects.\\

{\footnotesize
 \firsttabfalse
\begin{table}[h]
\[
 \begin{array}{|l|l|l|l|}
 \hline \hline
Process & Experiment\, [Ref.] & Observable & \# Data \\ \hline \hline
   \mbox{DIS} & BCDMS~\cite{BCDMS}  &F^p_2, F^d_2 &(177,159) \\
\cline{2-4}
& E665~\cite{E665} &F^p_2, F^d_2 &(53,53) \\ \cline{2-4}
& H1~\cite{H1}, ZEUS~\cite{ZEUS} &F^p_2 &(150,158) \\ \cline{2-4}
& NMC~\cite{NMC97} &F^p_2, F^d_2 &(130,130)\\
 \cline{2-4}
& NMC~\cite{NMC94} &F^p_2-F^n_2 & (12)\\ 
  \hline
\mbox{SIDIS}        & HERMES~\cite{Hermes}&  \frac{\overline{d}-\overline{u}}{u-d}& (5) \\ \hline
\mbox{Drell-Yan}        & E866~\cite{E866n}&  \frac{\sigma_{pd}}{2\,\sigma_{pp}}& (15) \\ \hline
W^+\,W^- \mbox{Asymmetry}
& CDF~\cite{CDF} &\frac{\sigma_{\ell^+} - \sigma_{\ell^-}}{\sigma_{\ell^+} +
 \sigma_{\ell^-}}&(11) \\ \hline \hline
 \end{array}
 \]
\caption{Sets of data.}
 \end{table}
 }

Notice that in addition to NMC data on the proton and the deuteron structure functions, $F^p_2$ and $F^d_2$ respectively, we have also included the data on the difference $F^p_2-F^n_2$ as reported in reference \cite{NMC94}, which we take as an alternative indicator of the flavour asymmetry, in the same footing as E866 or HERMES data.

\section{Scenarios}

As it was stated above, it is clear that the last four sets of data in Table 1 have very little statistical weight relative to DIS data. In order to preserve the information 
contained in them, and highlight the alternative scenarios that they imply, we give them an artificial extra weight in the $\chi^2$ minimization procedure. For E866, NMC data on the difference $F^p_2-F^n_2$ and HERMES, these weights are varied  according to which experiment we want to particularize, yielding three distinctive parton distribution sets labeled Set I, Set II, and Set III, respectively. For comparison we also include a set in which no additional weights are given, labeled as Set 0.

In Set I, a moderate weight factor ($\omega_{E866}=10$) is applied to E866 data, allowing an excellent matching ($\chi^2_{E866}=13.4$) shown by the solid line in Figure 2b., a rather good agreement with HERMES ($\chi^2_{HERMES}=5.3$ Figure 2c.), while a rather poor accord with NMC ($\chi^2_{NMC}=153$ Figure 2a.).
This set has a close resemblance with MRST, and clearly favors the agreement with E866 data relative to that with NMC.  However, at variance with MRST, because it includes explicitely NMC data, it also reproduces it slightly better. It is also
very similar to the unweighted fit, labeled as Fit 0, due to the fact that E866
data has by itself the largest constraining power of the three sets of data.
In Table 2 we show the corresponding $\chi^2$ values obtained.

\pagebreak

\firstfigfalse
\setlength{\unitlength}{1.mm}
\begin{figure}[!hbt]
\begin{picture}(100,170)(0,0)
\put(-15,50){\mbox{\epsfxsize11.0cm\epsffile{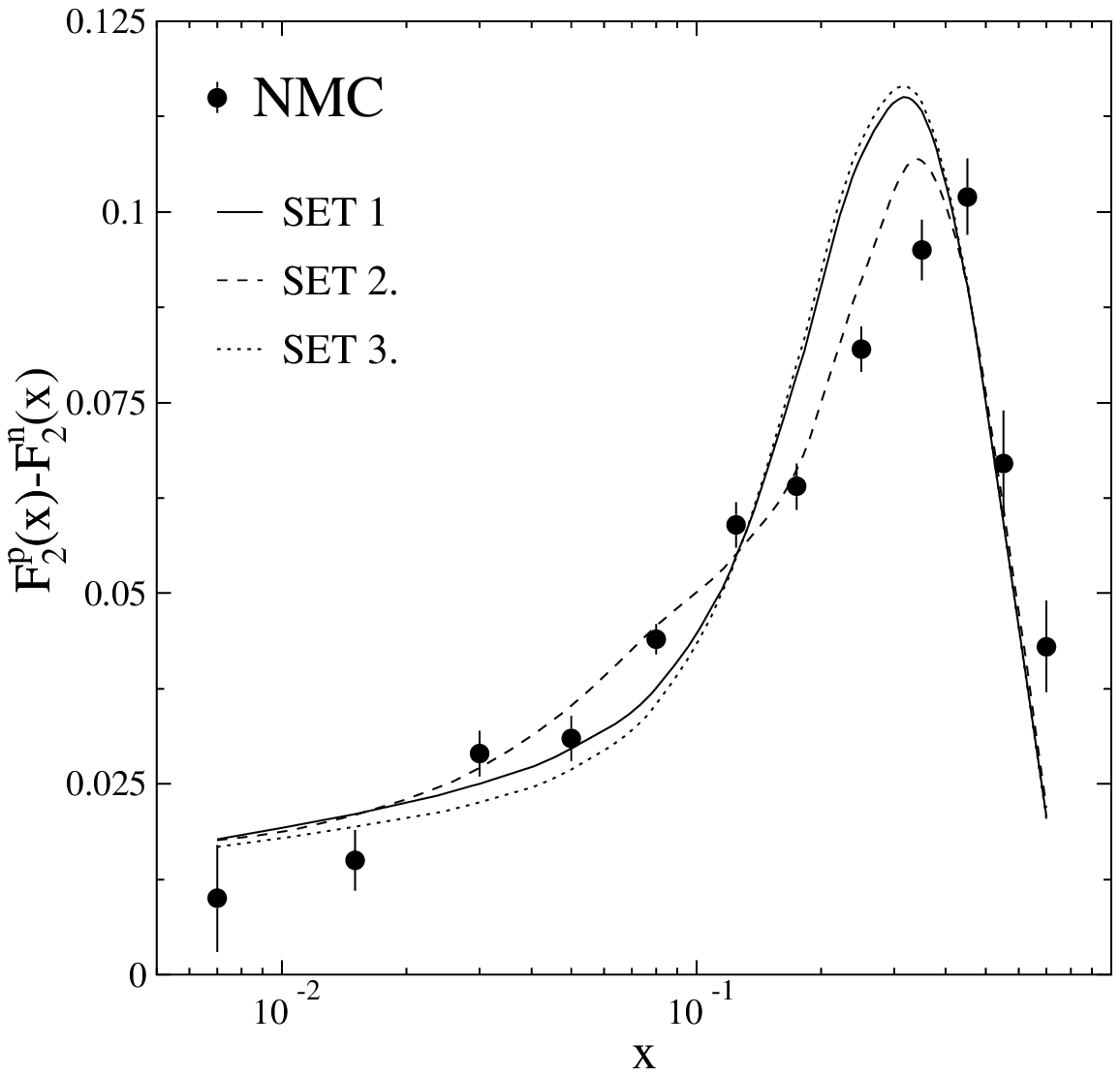}}}
\put(68,50){\mbox{\epsfxsize11.0cm\epsffile{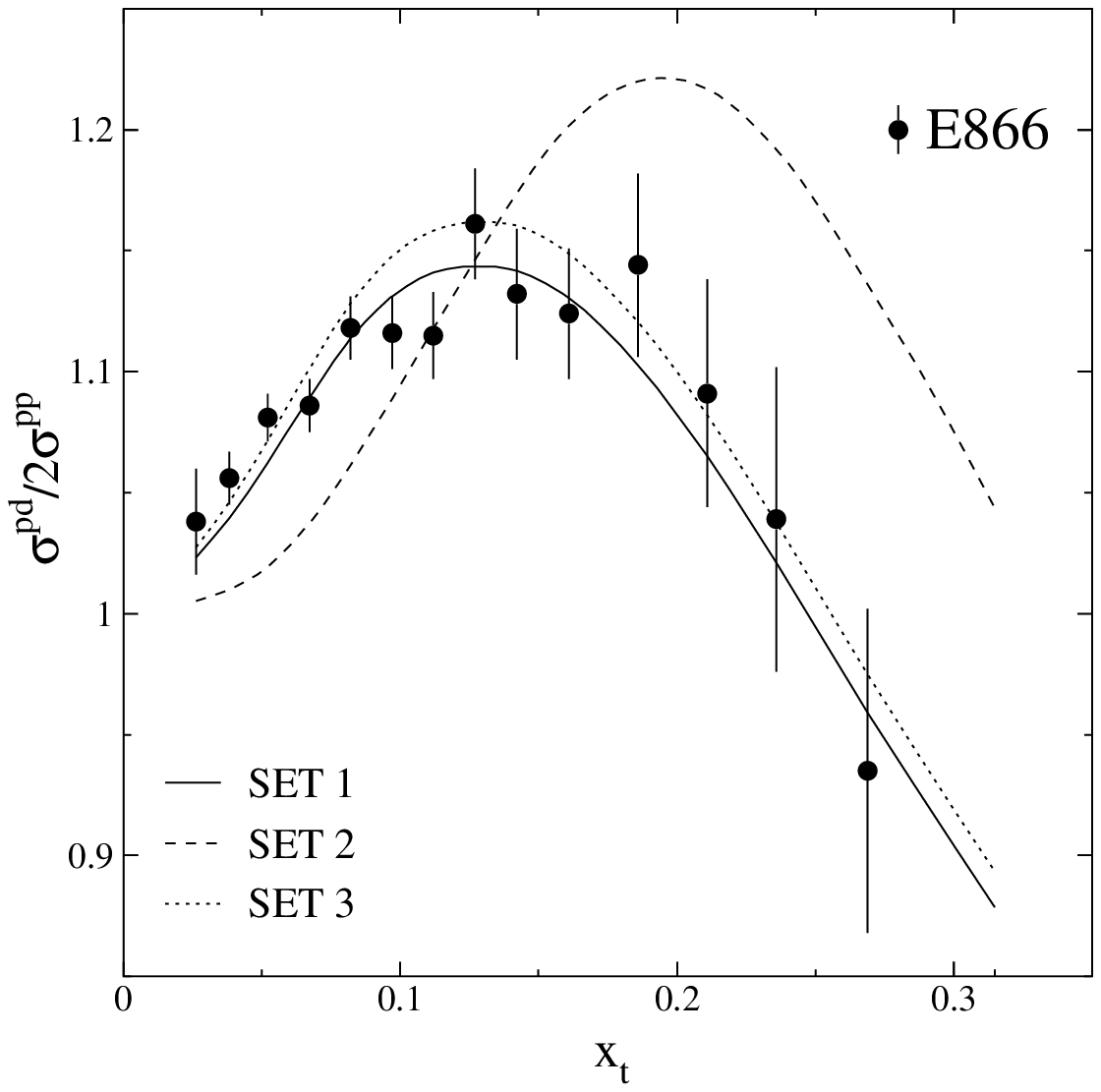}}}
\put(30,-32){\mbox{\epsfxsize11.0cm\epsffile{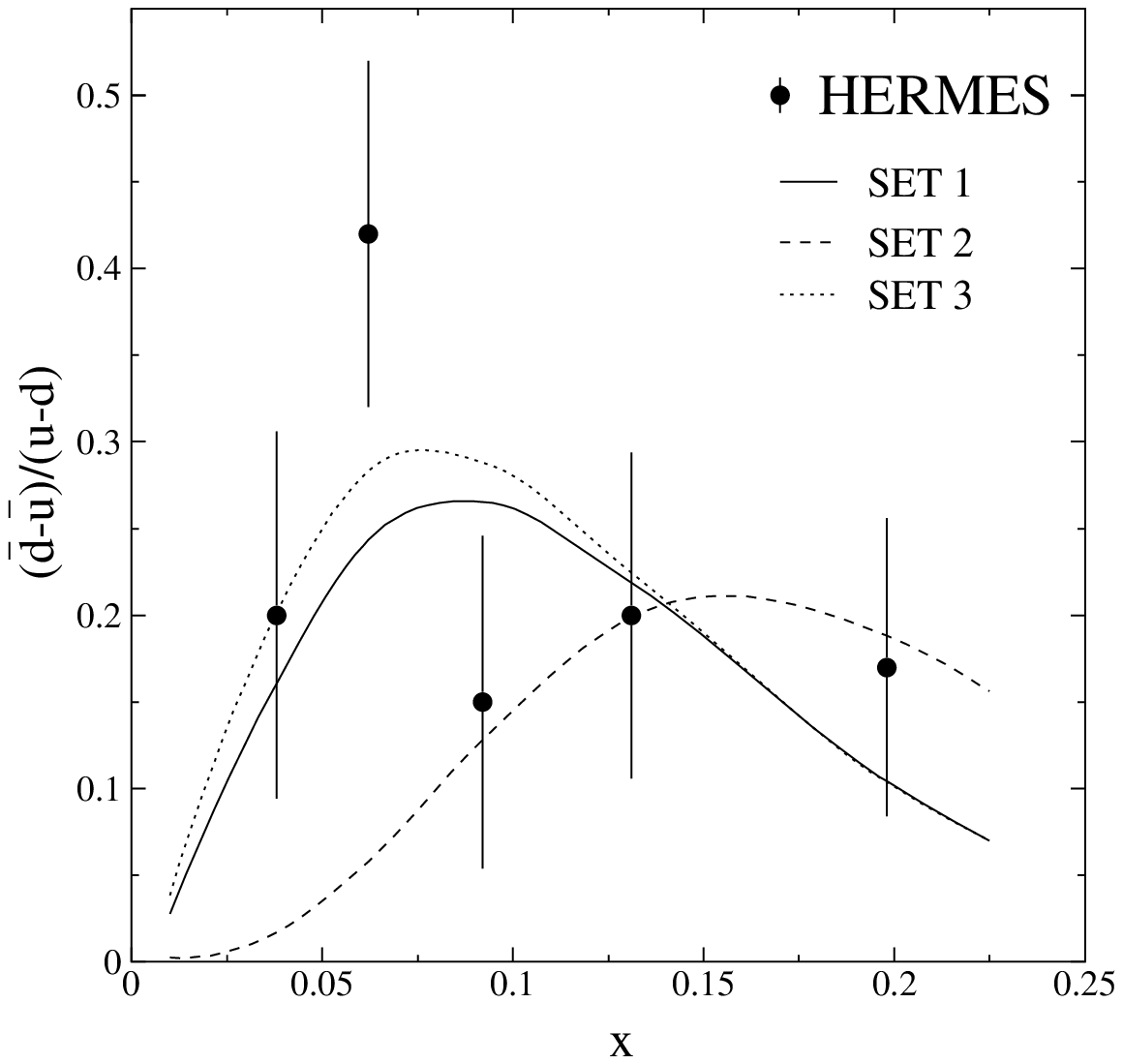}}}
\put(65,160){\mbox{\footnotesize \bf a)}}
\put(91,160){\mbox{\footnotesize \bf b)}}
\put(53,77){\mbox{\footnotesize \bf c)}}
\put(15,0){\mbox{\footnotesize \bf FIGURE 2:} 
{\footnotesize NMC (a), E886 (b), and HERMES (c) $\overline{d}/\overline{u}$-related measurements against fits.}} 
\end{picture}
\end{figure}

Set II, shown as dashes in Figure 2, have also a moderate weight factor on MNC data
($\omega_{NMC}=8$), which improves dramatically the agreement with this data ($\chi^2_{NMC}=45$), however E866 is not reproduced at all ($\chi^2_{E866}=142$). The agreement with HERMES is also significantly reduced ($\chi^2_{HERMES}=16$). Notice also that the overall quality of the fit deteriorates in a 10 \%.
 
Set III is designed to accommodate HERMES data ($\chi^2_{HERMES}=4.6$)
with a very strong  weight factor ($\omega_{HERMES}=180$), however it also reproduce fairly well E866 data ($\chi^2_{E866}=19$). The agreement with
NMC is completely lost ($\chi^2_{NMC}=184$), and it is even worse than in Set I.

It is interesting to notice that the most striking difference between the sea
quark densities inspired in the E866/HERMES and NMC scenarios, respectively, is 
that whereas NMC data favors $\overline{d}$ quark densities up to more than twice larger that $\overline{u}$ ones, both E866 and HERMES lead to not so large $\overline{d}/\overline{u}$ ratios, as shown in Figure 3a. There is also a clear difference in the value of momentum fraction at which the ratio reaches its maximum, which in turn determines the position of the peak of the $\overline{d}-\overline{u}$ distribution. Both HERMES and E886 lead to peaks around $x \sim 0.1$ while NMC favors $x \sim 0.2$ as can be seen in Figure 3b. 

\firstfigfalse
\setlength{\unitlength}{1.mm}
\begin{figure}[!hbt]
\begin{picture}(100,90)(0,0)
\put(-15,-32){\mbox{\epsfxsize11.0cm\epsffile{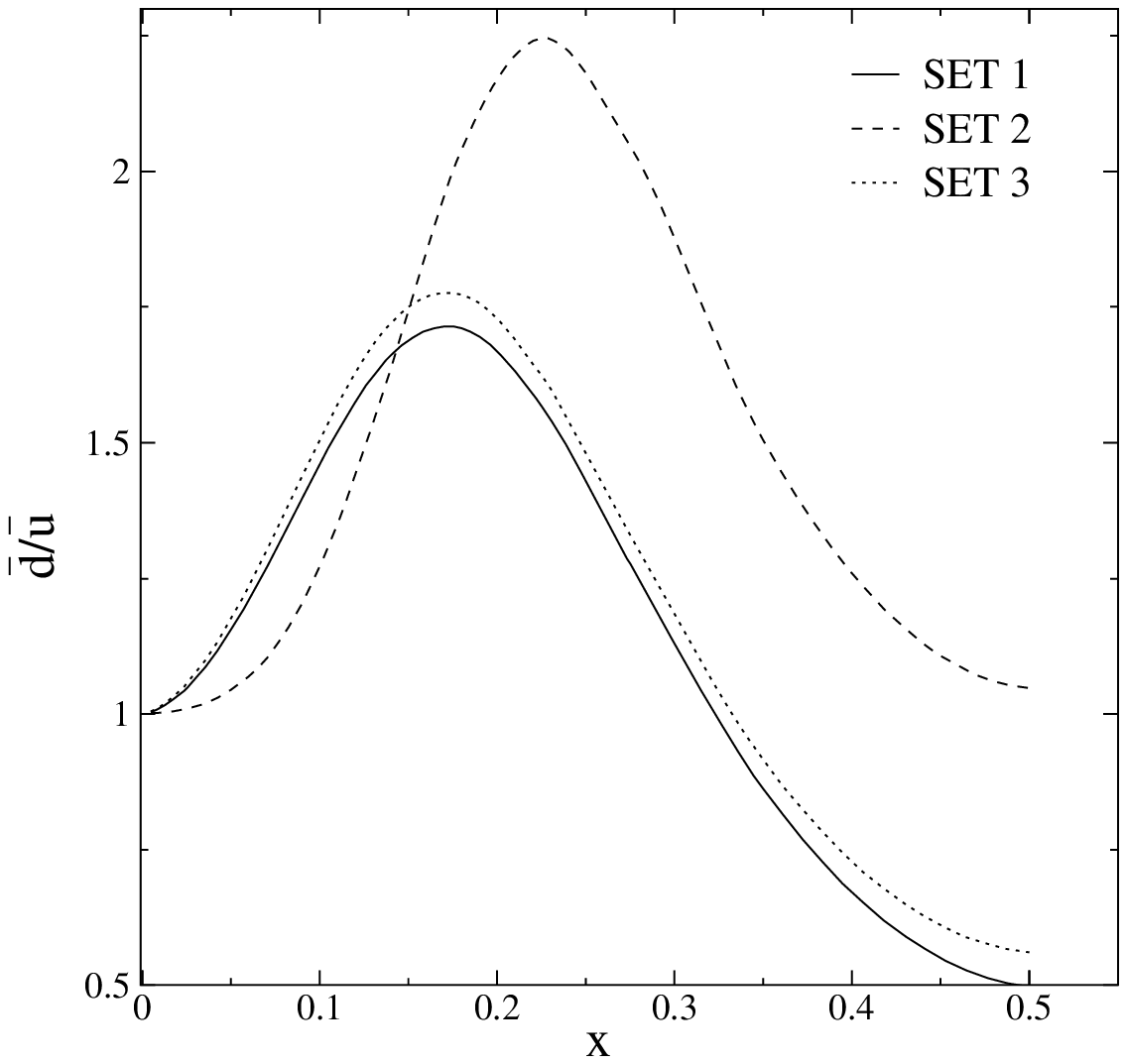}}}
\put(68,-32){\mbox{\epsfxsize11.0cm\epsffile{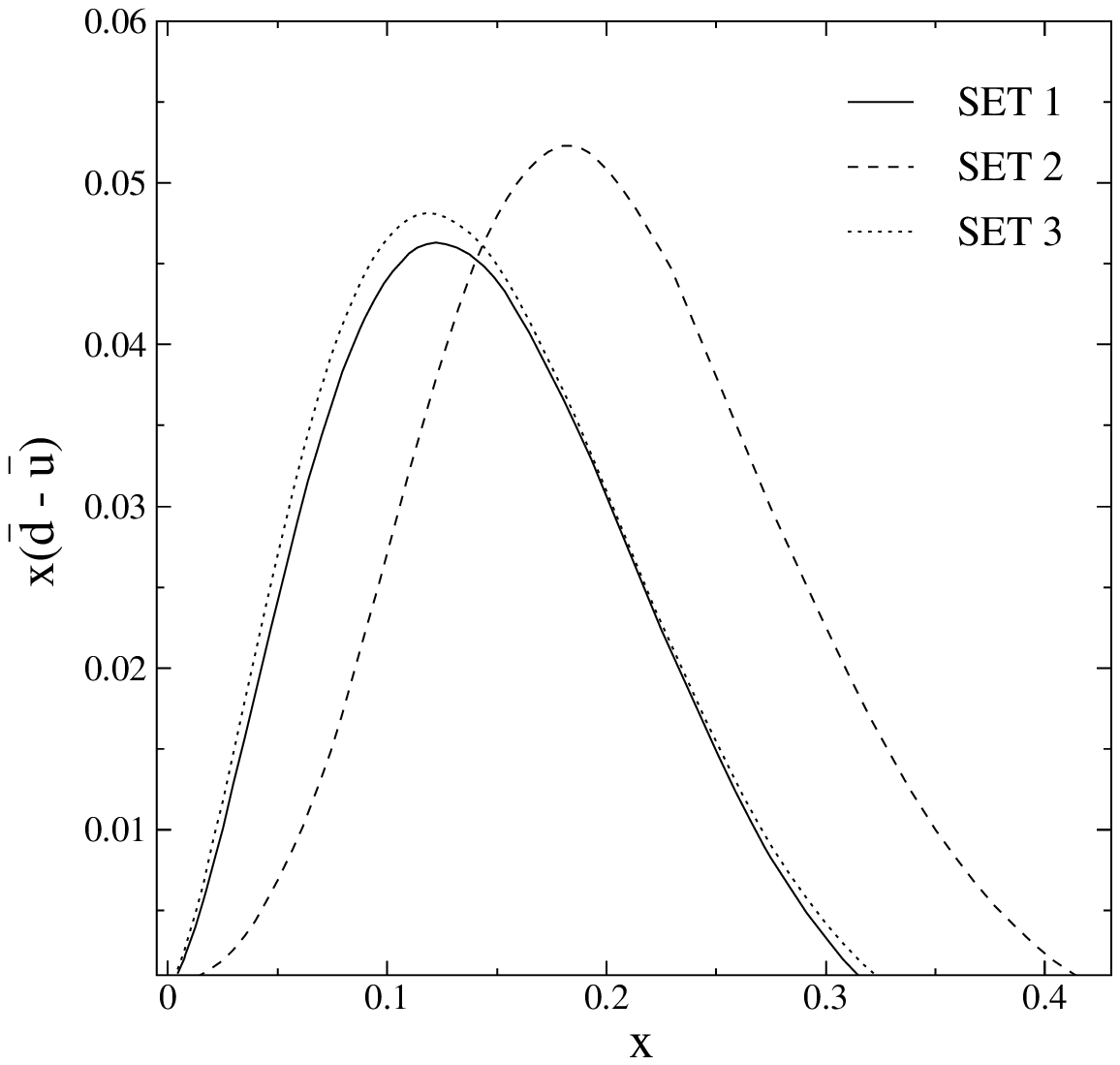}}}
\put(7,78){\mbox{\footnotesize \bf a)}} 
\put(90,78){\mbox{\footnotesize \bf b)}} 
\put(40,0){\mbox{\footnotesize \bf FIGURE 3:} 
{\footnotesize   {\bf a)} $\overline{d}/\overline{u}$  and {\bf b)} $\overline{d}-\overline{u}$ distributions from the fits.}} 
\end{picture}
\end{figure}

\firstfigfalse
\setlength{\unitlength}{1.mm}
\begin{figure}[!hbt]
\begin{picture}(100,90)(0,0)
\put(-15,-32){\mbox{\epsfxsize11.0cm\epsffile{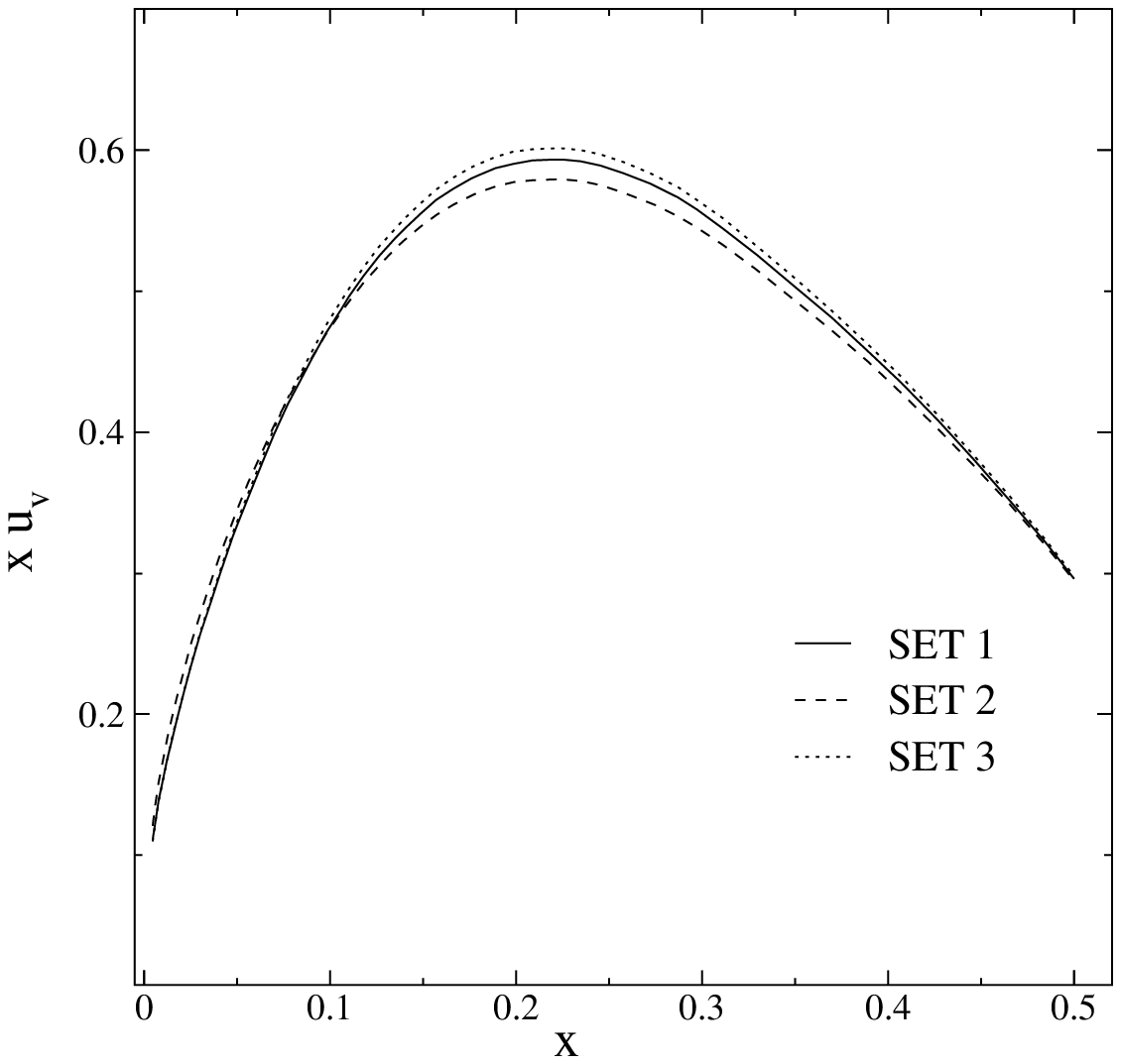}}}
\put(68,-32){\mbox{\epsfxsize11.0cm\epsffile{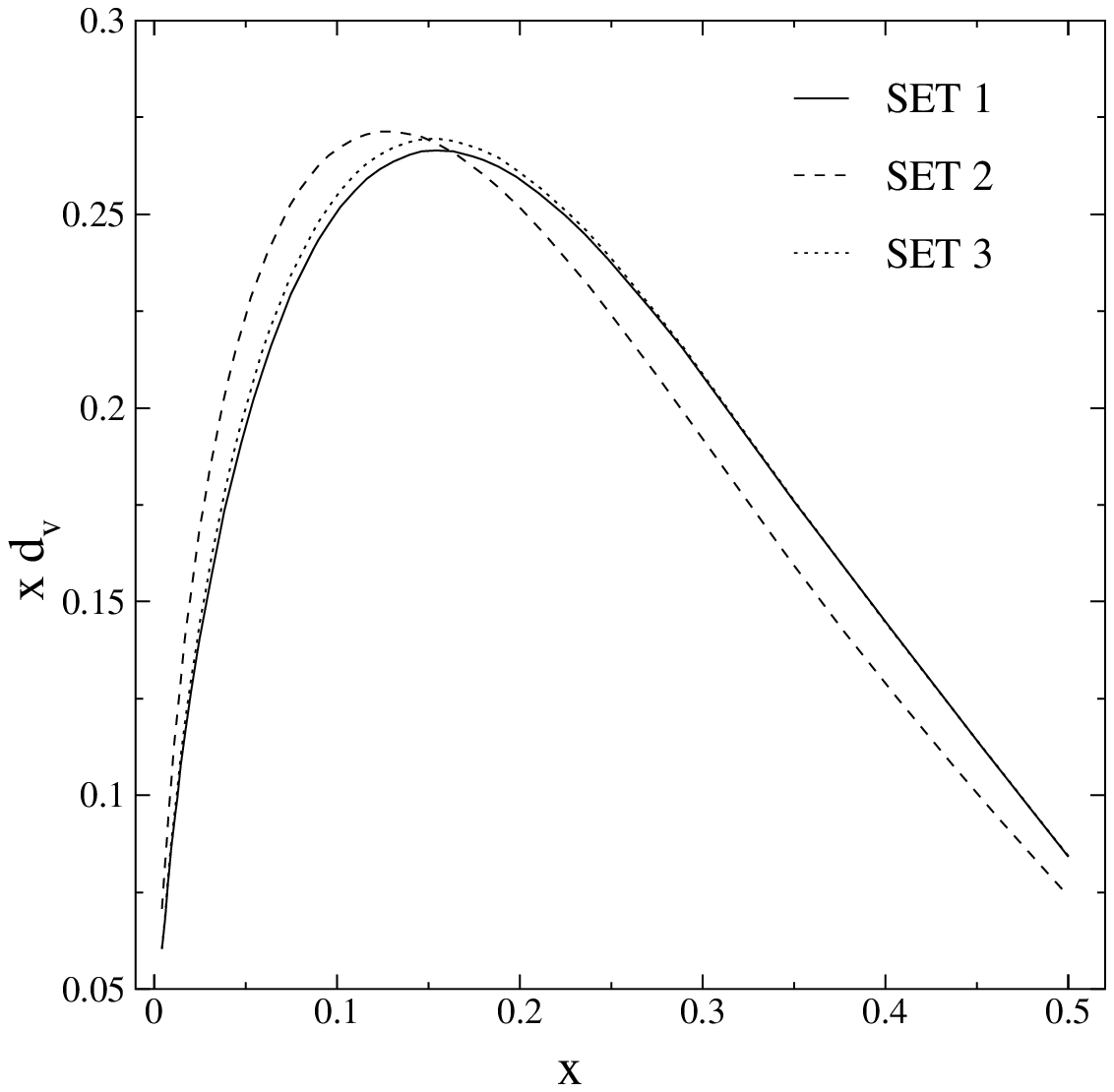}}}
\put(7,78){\mbox{\footnotesize \bf a)}} 
\put(90,78){\mbox{\footnotesize \bf b)}} \put(40,0){\mbox{\footnotesize \bf FIGURE 4:} 
{\footnotesize  Valence distributions from the fits.}} 
\end{picture}
\end{figure}

Given that neutral current structure function data are only sensitive to 
the sum of quark-antiquark densities of each flavour, the size of the flavour 
symmetry breaking in the sea introduces an additional uncertainty in the extraction of up and down valence quark
densities, as shown in Figures 4a and 4b, respectively.  There, it can be seen that again the NMC symmetry breaking scenario leads to 
an interesting difference in the behaviour of the valence densities, most noticeable in the case of the down quark distribution. \\  
 
{\footnotesize
 \firsttabfalse
\setcounter{table}{1}
 \begin{table}
 \[
 \begin{array}{|l|r|r|r|r|}
 \hline \hline
  & \mbox{I (E866)}  & \mbox{II (NMC)} & \mbox{III (Hermes)}& \mbox{0 ($\omega_i$)} \\
 \hline \hline
 NMC\,\, F_{2}^{p}&209.3&177.6&205.08&201.6\\
 NMC\,\, F_{2}^{D}&173.0&200.5&161.9&167.9\\
 BCDMS\,\, F_{2}^{p}&175.5&256.7&171.508&179.7\\
 BCDMS\,\, F_{2}^{D}&176.2&222.1&174.79&176.3\\
 E665\,\, F_{2}^{p}&61.9&62.9&61.8&62.4\\
 E665\,\, F_{2}^{D}&51.0&53.3&49.5&49.6\\
 H1\,\, F_{2}^{p}&114.6&118.7&127.4&116.5\\
 ZEUS\,\, F_{2}^{p}&243.2&239.5&246.5&239.9\\
 CDF\,\, A &33.9&12.5&34.7&27.9\\
 \hline
 E866\,\, \sigma^{pD}/2\sigma^{pp}&13.4&142.9&19.2&22.2\\
 NMC\,\, F_{2}^{n}-F_{2}^{p} &153.9&44.9&184.35&136.8\\
 HERMES\,\, \frac{\overline{d}-\overline{u}}{u-d} &5.3&16.2&4.6&6.5\\
 \hline \hline
 \chi^2 {\tiny}(unweighted)&1411.3& 1547.8&1442.1&1389.6\\
 \hline \hline
 \end{array} 
 \]
\caption{ {\footnotesize $\chi^2$ values obtained for different sets of data.}}
 \end{table}
 }

\section{Conclusions}

We have analysed the consequences of different measurements of the ratio $\overline{d}/\overline{u}$ asymmetry in the extraction of parton distributions.
We have found that even though E866 and HERMES results on $\overline{d}/\overline{u}$ lead to consistent extractions of parton distributions, NMC data on the difference of the proton and the neutron structure function can not be contained in this picture. 

In view of the next generation of high precision hadron collision experiments, 
the issue of uncertainties in parton distribution functions, and their precise origin, is a highly relevant one, that will certainly require further attention in the near future. 

Nuclear effects when using deuterium targets is a more or less obvious
source of uncertainties in the flavour asymmetry measurements we have deal with.
But at variance with the most conventional estimates for nuclear effects in deuterium, which lead to small corrections in the NMC observable at very small and large momentum fractions, the discrepancies between parton densities inspired in NMC data and those in agreement with E866 or HERMES data are concentrated at intermediate values of $x$.

These results suggest that even though the breakdown of the isospin symmetry in the sea of nucleons is indeed one of the reasons for the Gottfried sum rule deficit observed by NMC, it is far from being the answer to the question raised
almost ten years ago. 

\section{Acknowledgements}

This research was partially supported by CONICET, ANPCyT and Fundaci\'on Antorchas.

\section{Appendix}
{\footnotesize
 \firsttabfalse
\setcounter{table}{2}
 \begin{table}
 \[
 \begin{array}{|l|r|r|r|r|}
 \hline \hline
  & \mbox{I (E866)}  & \mbox{II (NMC)} & \mbox{III (Hermes)}& \mbox{0 ($\omega_i=0$)} \\
 \hline \hline
 \alpha_u &  0.305 &  0.372 &  0.314 &  0.311\\
 \beta_u  &  3.157 &  3.114 &  3.192 &  3.176\\
 \gamma_u & -0.132 & -0.356 & -0.138 & -0.002\\
 \delta_u &  16.84 &  10.07 &  16.94 &  17.12\\
 \alpha_d &  0.171 &  0.081 &  0.168 &  0.182\\
 \beta_d  &  3.598 &  3.394 &  3.550 &  3.639\\
 \gamma_d &  16.69 &  878.9 &  22.81 &  22.38\\
 \delta_d &  30.47 &  291.5 &  30.95 &  33.65\\
 \alpha_g &  0.467 &  0.778 &  0.670 &  0.583\\
 \beta_g  &  20.59 &  27.11 &  20.69 &  20.70\\
 \gamma_g & -3.502 & -2.130 & -4.165 & -3.467\\
 \delta_g &  8.084 &  15.35 &  8.577 &  7.877\\
 N_{\Sigma}& 0.484 &  0.659 &  0.421 &  0.423\\
 \alpha_{\Sigma} & -0.175 & -0.179 &  -0.210 & -0.210\\
 \beta_{\Sigma}  &  6.948 &  6.377 &  6.945 &  6.966\\
 \gamma_{\Sigma} & -1.086 & -3.806 & -1.012 & -1.066\\
 \delta_{\Sigma} &  11.22 &  13.29 &  11.45 &  11.32\\
 N_{\Delta}      &  5.545 &  41.85 &  4.832 &  5.557\\
 \alpha_{\Delta} &  1.826 &  3.885 &  1.729 &  1.913\\
 \beta_{\Delta}  &  10.83 &  20.41 &  10.94 &  10.86\\
 \gamma_{\Delta} &  11.74 &  274.1 &  11.31 &  12.26\\
 \delta_{\Delta} & -43.65 &  434.5 & -40.96 &  41.03\\
 \hline \hline
 \end{array} 
 \]
\caption{ {\footnotesize Parameters for the different sets.}}
 \end{table}
 }

\end{document}